\documentclass[journal]{IEEEtran}
\usepackage{amsfonts}
\usepackage{cite}
\usepackage{amsmath,minibox}
\usepackage{times,bbm}
\usepackage{graphicx}
\usepackage{amssymb}
\usepackage{caption}
\usepackage{color}
\usepackage{subcaption}
\usepackage{algorithm}
\usepackage{algorithmicx}
\usepackage{algpseudocode}
\usepackage[update,prepend]{epstopdf}

\allowdisplaybreaks

\begin{document}
\bstctlcite{IEEEexample:BSTcontrol}
\title{Localization of Multiple Targets with Identical Radar Signatures in Multipath Environments with Correlated Blocking}
\author{Sundar Aditya,~\IEEEmembership{Student~Member,~IEEE},  Andreas F. Molisch,~\IEEEmembership{Fellow,~IEEE}, Naif Rabeah,~\IEEEmembership{Member,~IEEE} and Hatim Behairy
\thanks{Sundar Aditya and Andreas F. Molisch are with the Wirless Devices and Systems (WiDeS) Lab, Ming Hsieh Dept. of Electrical Engineering, USC, Los Angeles, CA 90089, USA (email:\{sundarad,molisch\}@usc.edu).}
\thanks{Naif Rabeah and Hatim Behairy are with the King Abdulaziz City for Science and Technology (KACST), P. O. Box 6086, Riyadh 11442, Saudi Arabia (email: \{nbnrabeah,hbehairy\}@kacst.edu/sa) \hfill Revised: \today.}
\thanks{This work was supported by KACST under grant number 33-878. Part of this work was presented at the International Conference on Communications (ICC) Workshop on Advances in Network Localization and Navigation (ANLN), held at London in June, 2015.}
}
\maketitle

\begin{abstract}
This paper addresses the problem of localizing an unknown number of targets, all having the same radar signature, by a distributed MIMO radar consisting of single antenna transmitters and receivers that cannot determine directions of departure and arrival. Furthermore, we consider the presence of multipath propagation, and the possible (correlated) blocking of the direct paths (going from the transmitter and reflecting off a target to the receiver). In its most general form, this problem can be cast as a Bayesian estimation problem where every multipath component is accounted for. However, when the environment map is unknown, this problem is ill-posed and hence, a tractable approximation is derived where only direct paths are accounted for. In particular, we take into account the correlated blocking by scatterers in the environment which appears as a prior term in the Bayesian estimation framework. A sub-optimal polynomial-time algorithm to solve the Bayesian multi-target localization problem with correlated blocking is proposed and its performance is evaluated using simulations. We found that when correlated blocking is severe, assuming the blocking events to be independent and having constant probability (as was done in previous papers) resulted in poor detection performance, with false alarms more likely to occur than detections.
\end{abstract}


\section{Introduction}
Over the past decade, there has been an increasing demand for accurate indoor localization solutions to support a multitude of applications, ranging from providing location-based advertising to users in a shopping mall \cite{Steiniger_et_al_2006}, to better solutions for tracking inventory in a warehouse \cite{select}, to assisted-living applications \cite{Witrisal_et_al_2016}. {\color{black}In many of these applications, the availability of location information enhances communication protocols (e.g., in a warehouse, communication with a tag can take place once it is within range of a reader) and the importance of localization to such communications systems has led to standardization efforts such as IEEE 802.15.4a \cite{ieee_802.15.4a}, which is a standard for joint localization and communication.} While localization of cellular devices falls under the category of  active localization (where the target transmits a ranging signal), the current paper will focus on the important case of passive localization, where the target only reflects radio-frequency (RF) signals (i.e., radar). In the above-mentioned applications, there are usually multiple targets present that have nearly identical radar signatures and hence, cannot be distinguished on that basis alone.

The typical localization architecture involves the deployment of multiple transmitters (TXs) and receivers (RXs) and hence, can be modeled using the distributed MIMO (multiple-input multiple-output) radar framework \cite{haimovich_blum_et_al_2008}. Due to cost and space constraints (e.g., in sensor applications), each TX and RX may be equipped with a single antenna only\footnote{\color{black}Since a major contributor to the cost is the amplifier in a RF chain, a cheaper distributed MIMO architecture can be realized by deploying a single RF chain each for the TX and RX antennas and switching between them.}. Hence, the radar cannot exploit the information contained in the angles of arrival/departure from the target-reflected signal as direction finding requires not just multiple antennas, but also careful calibration of the antenna elements, which might be cost-prohibitive in the case of sensor nodes and similar devices. This motivates the study of multi-target localization (MTL) without angular information in an indoor distributed MIMO radar setting.

In such a setting, a direct path (DP) is one that propagates directly from TX to target to RX. Each DP gives rise to an ellipse-shaped ambiguity region passing through the target location, with the TX and RX at the foci, and the intersection of three or more such curves unambiguously determine the target location. For indoor localization, the following additional challenges arise - (i) targets can be blocked by non-target scatterers such as walls, furniture etc., (ii) the scatterers can also give rise to indirect paths (IPs) which need to be distinguished from DPs, (iii) in the presence of multiple targets, yet another challenge is to \emph{match} the DPs to the right targets\footnote{This process is also referred to as data association \cite{bar1995multitarget, Mah_2007}. Throughout this work, we shall use the term matching to refer to data association.}. An incorrect matching would result in ghost targets \cite{Shen_and_Molisch_2013_2}.

\begin{figure}
 \centering
 \includegraphics[scale=0.25]{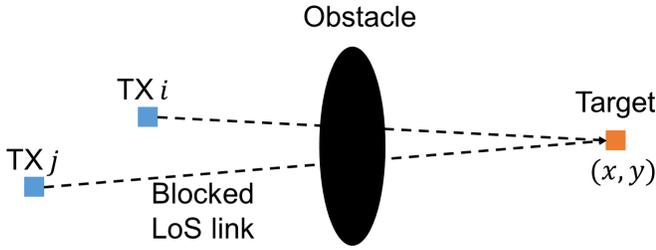}
 \caption{Correlated blocking: An example}
 \label{fig:1} 
\end{figure}

There has been a considerable amount of literature on MIMO radar over the last decade. The fundamental limits of localization in MIMO radar networks were studied in \cite{Godrich_et_al_2008}. A number of works have dealt with MTL using co-located antenna arrays at the TX and RX \cite{Zheng_et_al_2012, li.etal_2013, yan.etal_2008, Liu_2010, Gorji_2010_et_al, Duofang_et_al_2008, Jinli_et_al_2008, Jin_et_al_2009, Chen_et_al_2010, Miao_et_al_2008, Gorji_et_al_2012, xia.etal_2007, xu.li_2007}. The single-target localization problem using widely-spaced antenna arrays was investigated in \cite{He_et_al_2010, Niu_et_al_2012} and the multi-target case in \cite{YueAi_2014}. None of these works address the issues of blocking and multipath common in an indoor environment. The works closest to ours are \cite{Kim_Lee_2014} and \cite{Shen_and_Molisch_2013_2}, where MTL in a distributed MIMO radar setting is addressed. The experiments and the system model in \cite{Kim_Lee_2014} do not consider the effect of blocking and a brute-force method is used for matching the DPs to targets, which is computationally infeasible for a large number of targets, as shown in Section~\ref{sec:algo}. On the other hand,\cite{Shen_and_Molisch_2013_2} considers the effect of blocking, but relies on the assumption of a constant and independent blocking probability for all DPs. In reality however, the DP blocking events in any environment are not mutually independent. As shown in Fig. \ref{fig:1}, the location of the two TXs is such that if one of them has line-of-sight (LoS) to the target, it is highly \emph{likely} that the other does as well. Similarly, if one of them is blocked with respect to the target, it is highly \emph{likely} that the other is as well. In other words, the DP blocking events are, in general, correlated and the extent of correlation depends on the network geometry. In this work, we investigate how correlated blocking can be exploited to obtain better location estimates for the targets.

Intuitively, our approach works as follows: when three or more ellipses intersect at a point, we first assume that they are DPs. We then compute the joint probability that LoS exists to the TXs and RXs in question at the point of intersection. If the probability is sufficiently high, then we conclude that a target is present.

The main contributions of this work are as follows:
\begin{itemize}
 \item The general problem of localizing all targets and scatterers in an unknown environment is cast as a Bayesian estimation problem. Such a fundamental formulation of the problem goes beyond the description in \cite{Shen_and_Molisch_2013_2} (and other papers, to the best of our knowledge).
 \item We show this problem to be ill-posed and proceed to derive a tractable approximation called the Bayesian MTL problem, where the objective is to localize only the targets, but not the scatterers. This is also a Bayesian estimation problem where the joint DP blocking distribution plays the role of a prior.
 \item We propose a polynomial time approximate algorithm to solve the Bayesian MTL problem, which can be used even when only empirical blocking statistics, obtained via measurements or simulations, are available.
\end{itemize} 

This paper consists of six sections. In the system model in Section \ref{sec:sysmodel}, we define decision variables to decide if a multipath component (MPC) is a DP, IP or a noise peak. The generalized problem of localizing all targets and scatterers is formulated as a Bayesian estimation problem in Section \ref{sec:BLP}, where along with the target and scatterer locations, the aforementioned decision variables are the estimation parameters. Furthermore, this problem is shown to be ill-posed and a more tractable approximation called the Bayesian MTL problem is derived, where the objective is to localize only the targets. In Section \ref{sec:algo}, the brute force solution to the Bayesian MTL problem is shown to have exponential complexity in the number of targets and TX-RX pairs (TRPs). As a result, a sub-optimal polynomial time algorithm taking correlated blocking into account is proposed instead. Simulation as well as experimental results for the proposed algorithm are presented in Section \ref{sec:simres} and finally, Section \ref{sec:summary} concludes the paper. 

\textbf{Notation:} Vectors and scalars are represented using bold (e.g., $\mathbf{x}$) and regular (e.g., $x$) lowercase letters, respectively. In particular, $\mathbf{1}$ denotes the all-one vector. For a collection of scalars $\{ a_{ij} : i\in J_1, j \in J_2 \}$, where $J_1$ and $J_2$ are discrete index sets, $vec(a_{ij})$ denotes the column vector containing all $a_{ij}$, ordered first according to index $i$, followed by $j$ and so on. Similarly, $\sum\limits_{i,j}$ and $\prod\limits_{i,j}$ respectively denote summation and product over index $i$, followed by $j$ and so on. For positive integers $a$ and $b$, $a \mbox{ \rm mod } b$ denotes the modulo operator, i.e., the remainder of the operation $a/b$. The set of real numbers is denoted by $\mathbb{R}$. For $x\in \mathbb{R}$, $\lfloor x \rfloor$ denotes the greatest integer lesser than or equal to $x$. For continuous random variables $X$ and $Y$, $f(X,Y)$ denotes their joint probability density function (pdf), $f(X)$ the marginal pdf of $X$, and $f(X|Y)$ the conditional pdf of $X$, given $Y$. $\mathbb{P}(.)$ and $\mathbb{E}[.]$ denote the probability and expectation operators, respectively.

\section{System Model}
\label{sec:sysmodel}
In this section, we introduce the formal notation to decide the identity of a MPC and model the correlated blocking of DPs. This lays the groundwork for the fundamental problem formulation that will then be solved in Sections \ref{sec:BLP} and \ref{sec:algo}.

Consider a distributed MIMO radar with $M_{\rm TX}$ TXs and $M_{\rm RX}$ RXs, each equipped with a single omni-directional antenna and deployed in an unknown environment. An unknown number of stationary point targets\footnote{The point target assumption simplifies the analysis of the problem, as each target can give rise to at most one DP. The impact of real (i.e., non-point) targets can be seen in our experimental results, presented in Section~\ref{sec:exp}.} are present and the objective is to localize all of them. We assume that the environment has non-target scatterers too, which can either block some target(s) to some TX(s) and/or RX(s), and/or give rise to IPs. All TX and RX locations are assumed to be known. The number of TRPs, denoted by $I$, equals $M_{\rm TX} M_{\rm RX}$. Unless otherwise mentioned, the convention throughout this work is that the $i$-th TRP $(i=1, \cdots, I)$ comprises of the $i_T$-th TX and $i_R$-th RX, where $i_T = 1+(i-1) \mbox{ \rm mod }M_{\rm TX}$ and $i_R = \lfloor (i-1)/M_{\rm TX} \rfloor + 1$ (Table \ref{tab:TRPeg}).
\begin{table}
 \centering
 \begin{tabular}{|c|c|c|c|c|c|c|c|c|c|}
  \hline
  TRP No. & 1 & 2 & 3 & 4 & 5 & 6 & 7 & 8 & 9 \\
  \hline
  TX No. & 1 & 2 & 3 & 1 & 2 & 3 & 1 & 2 & 3 \\
  \hline
  RX No. & 1 & 1 & 1 & 2 & 2 & 2 & 3 & 3 & 3 \\
  \hline
 \end{tabular}
 \caption{TRP indexing notation for the 3 TX, 3 RX case}
 \label{tab:TRPeg}
\end{table}

For ease of notation, we restrict our attention to the two-dimensional (2D) case, where all the TXs, RXs, scatterers and targets are assumed to lie in the horizontal plane. The extension to the 3D case is straightforward. Let the TX and the RX for the $i$-th TRP be located at $(c_i,d_i)$ and $(a_i,b_i)$, respectively. We assume that the TXs use orthogonal signals so that the RXs can distinguish between signals sent from different TXs. For each TRP, the RX extracts the channel impulse response from the measured receive signal; an MPC is assumed to exist at a particular delay (within the resolution limit of the RX) when the amplitude of the impulse response at that delay bin exceeds a threshold; alternately, a maximum likelihood (ML) estimator or other high-resolution algorithms can be used to extract the amplitudes and delays of all the MPCs \cite{CLEAN},\cite{Richter_2005}. All MPCs that do not involve a reflection off a target (e.g., TX$\rightarrow$scatterer$\rightarrow$RX) are assumed to be removed by a background cancellation technique. For stationary or even slow-moving targets, a simple way to achieve this is to measure the impulse responses for all the TRPs when no targets are present. This set of template signals can then be subtracted from the signals obtained when the target(s) are introduced, which would remove MPCs of the form TX$\rightarrow$scatterer$\rightarrow$RX since they appear twice \cite{salmi_molisch_2011}. Other background subtraction techniques for localization and tracking applications are described in \cite{Chen_Leung_Tian_2014},\cite{Zet_Esch_Jova_thoma_2015}. An MPC involving more than two reflections is assumed to be too weak to be detected. Finally, two or more MPCs could have their delays so close to one another that they can be unresolvable due to finite bandwidth. Under this model, each extracted MPC could be one or more of the following: 
\begin{itemize}
  \item[1.] A DP to one or more targets, which occurs when a target has LoS to both the TX and RX in question.
  \item[2.] An IP of the first kind, which is of the form TX$\rightarrow$target$\rightarrow$scatterer$\rightarrow$RX.
  \item[3.] An IP of the second kind, having the form TX$\rightarrow$scatterer$\rightarrow$target$\rightarrow$RX.
  \item[4.] A noise peak.
\end{itemize}

Each MPC gives rise to a time-of-arrival (ToA) estimate which, in turn, corresponds to a range estimate. If only additive white Gaussian noise (AWGN) is present at the RXs, then each ToA estimate is approximately perturbed by zero-mean Gaussian errors whose variance depends on the SNR via the Cramer-Rao lower bound (CRLB) and the choice of estimator \cite{Shen_et_al_2012}. For simplicity, it is assumed that all ToA estimation errors have the same variance $\hat{\sigma}^2$. The extension to the general case where the variance is different for each MPC is straightforward. Thus, for a DP, the true range of the target from its TRP is corrupted by AWGN of variance $\sigma^2=c^2\hat{\sigma}^2$, where $c$ is the speed of light in the environment.

Suppose the $i$-th TRP has $N_i$ MPCs extracted from its received signal. Let $r_{ij}$ denote the range of the $j$-th extracted MPC at the $i$-th TRP and let $\mathbf{r}_i=[r_{i1}, \hspace*{1mm} r_{i2}, \hspace*{1mm} \cdots, \hspace*{1mm} r_{iN_i}] \in \mathbb{R}^{N_i \times 1}$ denote the vector of range estimates from the $i$-th TRP. Similarly, let $\mathbf{r}=[\mathbf{r}_1, \hspace*{1mm} \mathbf{r}_2, \hspace*{1mm} \cdots, \hspace*{1mm} \mathbf{r}_I] \in \mathbb{R}^{N_1N_2...N_I \times 1}$ denote the stacked vector of range estimates from all TRPs. If $r_{ij}$ is a DP corresponding to a target at $(x_t,y_t)$, then the conditional pdf of $r_{ij}$, given $(x_t,y_t)$, is Gaussian and denoted by $f_{\rm DP}(r_{ij}|x_t,y_t)$ and has the following expression:
\begin{align}
 \label{eq:f_DP}
 f_{\rm DP}(r_{ij}|x_t,y_t) &= \frac{1}{\sqrt{2\pi}\sigma}\exp\left[-\frac{(r_{ij}- r_i(x_t,y_t))^2}{2\sigma^2}\right] \\
 \mbox{where,} \hspace{1mm} r_i(x_t,y_t) &= \sqrt{(x_t-a_i)^2+(y_t-b_i)^2} \notag \\
 & ~ + \sqrt{(x_t-c_i)^2+(y_t-d_i)^2} \notag
\end{align}
$r_i(x_t,y_t)$ denotes the range of a target at $(x_t,y_t)$ from the $i$-th TRP. Similarly, let $f_{\rm IP,1}(r_{ij}|x_t,y_t,u_m,v_m)$ and $f_{\rm IP,2}(r_{ij}|x_t,y_t,u_m,v_m)$ denote the conditional IP pdfs of the first and second kind, respectively, given a target at $(x_t,y_t)$ and a scatterer at $(u_m,v_m)$. These pdfs are also Gaussian,
\begin{align}
 f_{\rm IP,1}(r_{ij}|x_t,y_t,u_m,v_m) &= \notag \\
 \frac{1}{\sqrt{2\pi}\sigma}\exp&\left[-\frac{(r_{ij}- l_i(x_t,y_t,u_m,v_m))^2}{2\sigma^2}\right] \\
 f_{\rm IP,2}(r_{ij}|x_t,y_t,u_m,v_m) &= \notag \\
 \frac{1}{\sqrt{2\pi}\sigma}\exp&\left[-\frac{(r_{ij}- m_i(x_t,y_t,u_m,v_m))^2}{2\sigma^2}\right] \\
 \mbox{where}~  l_i(x_t,y_t,u_m,v_m) &= \sqrt{(c_i-x_t)^2+(d_i-y_t)^2} \notag \\
 +\sqrt{(x_t-u_m)^2+(y_t-v_m)^2} &+ \sqrt{(u_m-a_i)^2 +(v_m-b_i)^2} \notag\\
 m_i(x_t,y_t,u_m,v_m) &= \sqrt{(c_i-u_m)^2+(d_i-v_m)^2} \notag \\
 +\sqrt{(u_m-x_t)^2+(v_m-y_t)^2} &+ \sqrt{(x_t-a_i)^2+(y_t-b_i)^2}  \notag
\end{align}
$l_i(x_t,y_t,u_m,v_m)$ and $m_i(x_t,y_t,u_m,v_m)$ respectively denote the path length between the $i$-th TRP, a target at $(x_t,y_t)$ and a scatterer at $(u_m,v_m)$ for an IP of the first and second kind. Finally, the range of a noise peak is modelled as a uniform random variable in the interval $[0,R_{\rm obs}]$, where $R_{\rm obs}$ denotes the maximum observable range in the region of interest.

Let the number of targets and scatterers be denoted by $T$ and $M$, respectively. To determine all the unknowns, every MPC needs to be accounted for. Hence, we define the following variables, 
\begin{align}
 \label{eq:ki}
 k_{it}&=\begin{cases}
 			 1, &\mbox{if $t$-th target is NOT blocked to the $i$-th TRP} \\
 			 0, &\mbox{else}
 		   \end{cases}\\
\label{eq:gij} 		   
 g_{imt}&=\begin{cases}
 			1, &\minibox{if $\exists$ an IP of the first kind between the $i$-th 			\\ TRP, $m$-th scatterer and $t$-th target} \\
 			0, &\mbox{else}
 		   \end{cases}\\
\label{eq:hij} 		   
 h_{imt}&=\begin{cases}
 			1, &\minibox{if $\exists$ an IP of the second kind between the $i$-th \\ TRP, $m$-th scatterer and $t$-th target}\\
 			0, &\mbox{else}
 		   \end{cases} 		    		   
\end{align}
The values of $k_{it}$, $g_{imt}$ and $h_{imt}$ ($i\in \{1, \cdots, I\}$, $m\in\{1, \cdots, M\}$, $t\in\{1, \cdots, T\}$) capture the ground truth regarding the existence of DPs and IPs and depend on the map of the environment, which is unknown. Therefore, these quantities need to be estimated from $\mathbf{r}$. To do this, we define the following decision variables to determine if an MPC $r_{ij}$ is a DP, IP or noise peak,
\begin{align}
 \tilde{k}_{ijt} &= \begin{cases}
 				 1, &\mbox{if $r_{ij}$ is a DP to the $t$-th target} \\
 				 0, &\mbox{else}
 				\end{cases}\\
 \tilde{g}_{ijmt} &= \begin{cases}
 				 1, &\minibox{if $r_{ij}$ is an IP of the first kind between the \\ $m$-th scatterer and $t$-th target} \\
 				 0, &\mbox{else}
 				\end{cases}\\
 \tilde{h}_{ijmt} &= \begin{cases}
 				 1, &\minibox{if $r_{ij}$ is an IP of the second kind between the \\ $m$-th scatterer and $t$-th target} \\
 				 0, &\mbox{else}
 				\end{cases}\\
 \tilde{z}_{ij}	&= \begin{cases}
 				 1, &\mbox{if $r_{ij}$ is a noise peak} \\
 				 0, &\mbox{else}
 				\end{cases} 				
\end{align}
Since two or more resolvable MPCs cannot be DPs to the same target or IPs of a particular kind between a given target-scatterer pair, it follows that the estimates of $k_{it}$, $g_{imt}$ and $h_{imt}$, denoted by $\hat{k}_{it}$, $\hat{g}_{imt}$ and $\hat{h}_{imt}$, respectively, are given by:
\begin{align}
\label{eq:constraints_k}
 \hat{k}_{it}=\displaystyle\sum\limits_{j=1}^{N_i} \tilde{k}_{ijt} \\
 \label{eq:constraints_g}
 \hat{g}_{imt}=\displaystyle\sum\limits_{j=1}^{N_i} \tilde{g}_{ijmt} \\
 \label{eq:constraints_h}
 \hat{h}_{imt}=\displaystyle\sum\limits_{j=1}^{N_i} \tilde{h}_{ijmt} 
\end{align}
Before concluding this section, we define the following vectors which shall be useful when the Bayesian MTL problem is defined in the next section
\begin{align}
\label{eq:reqvecs}
 \mbox{Ground truth:}~ \mathbf{k} &= vec(k_{it})\\
  \mathbf{g} &= vec(g_{imt}) \\
  \mathbf{h} &= vec(h_{imt}) \\
 \mbox{DP/IP/noise peak decisions:}~ \tilde{\mathbf{k}} &= vec(\tilde{k}_{ijt})\\
 \tilde{\mathbf{g}}&= vec(\tilde{g}_{ijmt})\\ \tilde{\mathbf{h}} &= vec(\tilde{h}_{ijmt})\\\tilde{\mathbf{z}} &= vec(\tilde{z}_{ij}) \\
 \mbox{Estimates of ground truth:}~ \hat{\mathbf{k}}&=vec(\hat{k}_{it}) \\\hat{\mathbf{g}}&=vec(\hat{g}_{imt}) \\\hat{\mathbf{h}}&=vec(\hat{h}_{imt})
\end{align}

\section{Bayesian MTL}
\label{sec:BLP}
Using the notation from the previous section, the MTL problem in multipath environments with correlated blocking is formulated as a Bayesian estimation problem in this section. We first show that the scatterer locations cannot be determined uniquely, in general, as they are not point objects. Then, we show that that the distribution of $\mathbf{k}$ in (\ref{eq:reqvecs}) captures correlated blocking in its entirety and acts as a prior. We also assume a single error at most between the entries of $\hat{\mathbf{k}}_t$ and $\mathbf{k}_t$ in order to obtain a tractable algorithm for the MTL problem in Section~\ref{sec:algo}.

Let $\Theta_{\rm tar}=\{(x_t,y_t):t=1,\cdots,T\}$ and $\Theta_{\rm sc}=\{(u_m,v_m):1,\cdots,M\}$ denote the collection of target and scatterer locations, respectively, and let $\tilde{\mathbf{p}}_{\rm dec}=[\tilde{\mathbf{k}},\tilde{\mathbf{g}},\tilde{\mathbf{h}}, \tilde{\mathbf{z}}]$ denote the vector of decision variables. Using the terminology defined in Section \ref{sec:sysmodel}, determining the location of all targets and scatterers can be formulated as a Bayesian estimation problem in the following manner,
\begin{align}
 \underset{\substack{T, M, \Theta_{\rm tar}, \Theta_{\rm sc}, \\\tilde{\mathbf{p}}_{\rm dec}, \mathbf{k},\mathbf{g},\mathbf{h}}} {\mbox{maximize}} ~  f(\mathbf{r}|\tilde{\mathbf{p}}_{\rm dec},\Theta_{\rm tar}&,\Theta_{\rm sc},\mathbf{k},\mathbf{g},\mathbf{h}) \times f(\Theta_{\rm tar},\Theta_{\rm sc}) \notag \\
\label{eq:Bayes} 
  \times \mathbb{P}(\hat{\mathbf{k}},\hat{\mathbf{g}},\hat{\mathbf{h}}|\Theta_{\rm tar},\Theta_{\rm sc}&,\mathbf{k},\mathbf{g},\mathbf{h}) \times \mathbb{P}(\mathbf{k},\mathbf{g},\mathbf{h}|\Theta_{\rm tar},\Theta_{\rm sc}) \\
 \mbox{subject to}~  (\ref{eq:constraints_k}), (\ref{eq:constraints_g}), (\ref{eq:constraints_h})&  \notag \\
  \label{eq:con2}
\sum_{j,t} \tilde{k}_{ijt} + \sum_{j,m,t} (\tilde{g}_{ijmt}+&\tilde{h}_{ijmt})+ \sum_{j} \tilde{z}_{ij} \geq N_i , ~ \forall i \\
 \label{eq:con3}
\tilde{k}_{ijt}, \tilde{g}_{ijmt}, \tilde{h}_{ijmt}, \hat{k}_{it}, \hat{g}_{imt}, &\hat{h}_{imt} \in \{0,1\}, ~ \forall i,j,t,m
\end{align}
where, the first term in the objective (\ref{eq:Bayes}) denotes the likelihood function and the remaining three terms denote the prior. A detailed explanation of all the terms and constraints in (\ref{eq:Bayes})-(\ref{eq:con3}) is provided below:

\begin{itemize}
 \item[(a)] The term $f(\Theta_{\rm tar},\Theta_{\rm sc})$ denotes the \emph{prior} joint distribution of the target and scatterer locations. It is reasonable to assume that the target and scatterer locations are independent of each other. Hence, $f(\Theta_{\rm tar},\Theta_{\rm sc})=f(\Theta_{\rm tar})f(\Theta_{\rm sc})$. In addition, $f(\Theta_{\rm tar})$ and $f(\Theta_{\rm sc})$ are both assumed to be uniform pdfs over the region of interest.
 \item[(b)] The discrete distribution $\mathbb{P}(\mathbf{k},\mathbf{g},\mathbf{h}|\Theta_{\rm tar},\Theta_{\rm sc})$ represents the geometry of the environment, such as the blocked DPs for each TRP, the IPs (if any) between a target-scatterer pair etc. Let $\Theta_{\rm TX}=\{(c_i,d_i):i=1,\cdots,I\}$ and $\Theta_{\rm RX}=\{(a_i,b_i):i=1,\cdots,I\}$ denote the collection of TX and RX locations, respectively. $\Theta_{\rm TX}$ and $\Theta_{\rm RX}$ are known quantities and for a given set of values for $\Theta_{\rm tar}$ and $\Theta_{\rm sc}$, the set $\Theta_{\rm env}=\{\Theta_{\rm TX}, \Theta_{\rm RX}, \Theta_{\rm tar}, \Theta_{\rm sc}\}$ completely describes  all the propagation paths in the environment and the values of $\mathbf{k}$, $\mathbf{g}$ and $\mathbf{h}$ are deterministic functions of $\Theta_{\rm env}$, denoted by $\mathbf{k}^{\rm(det)}(\Theta_{\rm env})$, $\mathbf{g}^{\rm(det)}(\Theta_{\rm env})$ and $\mathbf{h}^{\rm(det)}(\Theta_{\rm env})$, respectively\footnote{This is akin to ray-tracing}. Hence,
 \begin{align}
 \label{eq:env_det}
  \mathbb{P}(\mathbf{k},\mathbf{g},\mathbf{h}|\Theta_{\rm tar},\Theta_{\rm sc}) &= \mathbbm{1}_{\mathbf{k}^{\rm (det)}(\Theta_{\rm env})}(\mathbf{k})\times \mathbbm{1}_{\mathbf{g}^{\rm (det)}(\Theta_{\rm env})}(\mathbf{g}) \notag \\ &\hspace{5mm}\times\mathbbm{1}_{\mathbf{h}^{\rm (det)}(\Theta_{\rm env})}(\mathbf{h})
 \end{align}
where $\mathbbm{1}_{\mathbf{y}}(\mathbf{x})$ equals 1 if $\mathbf{x}=\mathbf{y}$ and 0, otherwise.

 \item[(c)] The estimates $\hat{k}_{it}$, $\hat{g}_{imt}$ and $\hat{h}_{imt}$ may differ from their respective ground truths, $k_{it}$, $g_{imt}$ and $h_{imt}$ due to noise or IPs. Assuming that $\hat{k}_{it}$ (or $\hat{g}_{imt}$, $\hat{h}_{imt}$) is conditionally independent of other estimates, given $k_{it}$ (or $g_{imt}$, $h_{imt}$), we get
 \begin{align}
 \label{eq:env_det2}
&\hspace{5mm} \mathbb{P}(\hat{\mathbf{k}},\hat{\mathbf{g}},\hat{\mathbf{h}}|\Theta_{\rm tar},\Theta_{\rm sc},\mathbf{k},\mathbf{g},\mathbf{h}) \notag \\
&= \mathbb{P}(\hat{\mathbf{k}},\hat{\mathbf{g}},\hat{\mathbf{h}}|\mathbf{k}^{\rm (det)}(\Theta_{\rm env}), \mathbf{g}^{\rm (det)}(\Theta_{\rm env}), \mathbf{h}^{\rm (det)}(\Theta_{\rm env})) \\
&=  \displaystyle\prod\limits_{i,t,m} \mathbb{P}(\hat{k}_{it}|k^{\rm (det)}_{it}(\Theta_{\rm env})) \times \mathbb{P}(\hat{g}_{imt}|g^{\rm (det)}_{imt}(\Theta_{\rm env})) \notag\\
&\hspace{25mm} \times \mathbb{P}(\hat{h}_{imt}|h^{\rm (det)}_{imt}(\Theta_{\rm env}))
 \end{align}
where (\ref{eq:env_det2}) follows from (\ref{eq:env_det}).
 
 \item[(d)] $\tilde{\mathbf{p}}_{\rm dec}$ is a sufficient statistic for estimating $\mathbf{k}$, $\mathbf{g}$ and $\mathbf{h}$. Hence, the likelihood function, $f(\mathbf{r}|\tilde{\mathbf{p}}_{\rm dec},\Theta_{\rm tar},\Theta_{\rm sc},\mathbf{k},\mathbf{g},\mathbf{h})$, equals $f(\mathbf{r}|\tilde{\mathbf{p}}_{\rm dec},\Theta_{\rm tar},\Theta_{\rm sc})$. Further, $f(\mathbf{r}|\tilde{\mathbf{p}}_{\rm dec},\Theta_{\rm tar},\Theta_{\rm sc})$ decomposes into product form as the noise terms on each $r_{ij}$ are mutually independent. Thus,
 \begin{align}
  f(\mathbf{r}|\tilde{\mathbf{p}}_{\rm dec},\Theta_{\rm tar},\Theta_{\rm sc},\mathbf{k},\mathbf{g},\mathbf{h}) &= f(\mathbf{r}|\tilde{\mathbf{p}}_{\rm dec},\Theta_{\rm tar},\Theta_{\rm sc}) \notag\\
&= \displaystyle\prod\limits_{i,j} f(r_{ij}|\tilde{\mathbf{p}}_{\rm dec},\Theta_{\rm tar},\Theta_{\rm sc}) \notag\\
\mbox{where,}~ f(r_{ij}|\tilde{\mathbf{p}}_{\rm dec},\Theta_{\rm tar},\Theta_{\rm sc}) &= \displaystyle\prod\limits_{t,m} (f_{\rm DP}(r_{ij}|x_t,y_t))^{\tilde{k}_{ijt}} \notag \\
\times (f_{\rm IP,1}&(r_{ij}|x_t,y_t,u_m,v_m))^{\tilde{g}_{ijmt}} \notag\\ 
 \times (f_{\rm IP,2}&(r_{ij}|x_t,y_t,u_m,v_m))^{\tilde{h}_{ijmt}} \notag\\
&\times \left(\frac{1}{R_{\rm obs}}\right)^{\tilde{z}_{ij}}
 \end{align}
  
 \item[(e)] Finally, constraint (\ref{eq:con2}) ensures that the number of DPs, IPs and noise peaks received at the $i$-th TRP is at least $N_i$, the number of resolvable MPCs extracted at the $i$-th TRP.
\end{itemize}

After taking natural logarithms, (\ref{eq:Bayes}) can be re-written as follows to obtain problem $P1$, where $\mathbf{k}$, $\mathbf{g}$ and $\mathbf{h}$ are no longer unknowns due to (\ref{eq:env_det}):
\begin{align}
 \label{eq:logBayes}
&P1:\underset{T,M,\tilde{\mathbf{p}}_{\rm dec},\Theta_{\rm tar},\Theta_{\rm sc}}{\mbox{minimize}} ~ \frac{1}{\sigma^2}\left[\displaystyle\sum\limits_{i,j,t} \tilde{k}_{ijt}(r_{ij} -r_i(x_t,y_t))^2 \right] \notag \\
& + \frac{1}{\sigma^2}\left[ \displaystyle\sum\limits_{i,j,t,m} \tilde{g}_{ijmt}(r_{ij}-l_i(x_t,y_t,u_m,v_m))^2 \right] \notag \\
& + \frac{1}{\sigma^2}\left[ \displaystyle\sum\limits_{i,j,t,m} \tilde{h}_{ijmt}(r_{ij}-m_i(x_t,y_t,u_m,v_m)^2\right]\notag\\
&+  \left[\displaystyle\sum\limits_{i,j,t} \tilde{k}_{ijt} + \displaystyle\sum\limits_{i,j,m,t} (\tilde{g}_{ijmt}+\tilde{h}_{ijmt})\right] \log\sqrt{2\pi}\sigma \notag \\
& + \left(\displaystyle\sum\limits_{i,j} \tilde{z}_{ij}\right) \log R_{\rm obs} - \sum_{i,t} \log\mathbb{P}(\hat{k}_{it}|k^{\rm (det)}_{it}(\Theta_{\rm env})) \notag\\  
& - \sum_{i,m,t} \log\mathbb{P}(\hat{g}_{imt}|g^{\rm (det)}_{imt}(\Theta_{\rm env})) \notag \\
&- \sum_{i,m,t} \log\mathbb{P}(\hat{h}_{imt}|h_{imt}^{\rm (det)}(\Theta_{\rm env}))  \\
& \mbox{subject to}~  (\ref{eq:constraints_k}), (\ref{eq:constraints_g}), (\ref{eq:constraints_h}), (\ref{eq:con2}), (\ref{eq:con3}) \notag
\end{align}

Typically, $\Theta_{\rm sc}$ represents a finite collection of points belonging to \emph{distributed} non-point objects (e.g., a wall), where reflection takes place. A minimum of three reflections are needed at each $(u_m,v_m)$ for uniquely determining $\Theta_{\rm sc}$, which need not be satisfied in all circumstances. Hence, $P1$ is ill-posed if the map of the environment is unknown\footnote{If the map of the environment is known, then $P1$ is not ill-posed and the IPs can be re-cast as \emph{virtual} DPs, obtained from virtual TXs and RXs \cite{setlur.etal_2012}.}. To make $P1$ tractable, we restrict ourselves to localizing only the targets by retaining those terms and constraints involving just the DPs in (\ref{eq:Bayes})-(\ref{eq:con3}). This gives rise to the following approximation, $P2$, which is also a Bayesian estimation problem that accounts for all the DPs
\begin{align}
  \label{eq:Bayesobj_approx1}
&P2:\underset{T,\tilde{\mathbf{k}},\Theta_{\rm tar}}{\mbox{minimize}} ~ \left(\displaystyle\sum\limits_{i,j,t} \tilde{k}_{ijt} \right)  \log\sqrt{2\pi}\sigma - \log \mathbb{P}(\mathbf{k}|\Theta_{\rm tar}) \notag \\
&+ \frac{1}{\sigma^2}\left[\displaystyle\sum\limits_{i,j,t} \tilde{k}_{ijt}(r_{ij} -r_i(x_t,y_t))^2 \right] - \sum_{i,t} \log\mathbb{P}(\hat{k}_{it}|k_{it}) \\
 \label{eq:con1_approx1}
& \mbox{subject to} ~ \tilde{k}_{ijt}, \hat{k}_{it} \in \{0,1\}, ~ \forall i,j,t,m \\
 \label{eq:con2_approx1}
 & \hspace{15mm}\sum_{j} \tilde{k}_{ijt} = \hat{k}_{it} 
\end{align}

The joint DP blocking distribution $\mathbb{P}(\mathbf{k}|\Theta_{\rm tar})$ in $P2$ is no longer a discrete-delta function, like (\ref{eq:env_det}). Instead, it depends on the distribution of scatterer locations in the environment. From (\ref{eq:ki}), $k_{it}=0$ if either the TX or the RX of the $i$-th TRP does not have LoS to $(x_t,y_t)$; hence, $k_{it}$ can be expressed as a product of two terms in the following manner:
 \begin{align}
  k_{it} &= v_{i_T,t} \times w_{i_R,t}, \\
 \mbox{where,}~  v_{i_T, t} &= \begin{cases}
   										& 1 , ~\mbox{if the $i_T$-th TX has LoS to $(x_t,y_t)$} \\
   										& 0 , ~\mbox{else}
  										\end{cases}\notag\\
   w_{i_R, t} &= \begin{cases}
   										& 1 , ~\mbox{if the $i_R$-th RX has LoS to $(x_t,y_t)$} \\
   										& 0 , ~\mbox{else}
  										\end{cases}\notag
 \end{align}
$k_{it}$ can be interpreted as a Bernoulli random variable when considering an ensemble of settings in which the scatterers are placed at random. For vectors $\mathbf{k}_t = [k_{1t}, \hspace{1mm}\cdots, \hspace{1mm} k_{It}]$, $\mathbf{v}_t = [v_{1,t}, \hspace{1mm}\cdots, \hspace{1mm} v_{M_{\rm TX}, t}]$ and $\mathbf{w}_t = [w_{1,t}, \hspace{1mm}\cdots, \hspace{1mm} w_{M_{\rm RX}, t}]$, it can be seen that $\mathbf{k}_t = \mathbf{w}_t \bigotimes \mathbf{v}_t$, where $\bigotimes$ denotes the Kronecker product. $\mathbf{k}_t$ is a vector of dependent Bernoulli random variables (Fig. \ref{fig:1}) and shall henceforth be referred to as the blocking vector at $(x_t,y_t)$. Note that $\mathbf{k}=vec(k_{it})=[\mathbf{k}_1, \hspace{1mm}\cdots, \hspace{1mm} \mathbf{k}_T]$ and therefore, $\mathbb{P}(\mathbf{k}|\Theta_{\rm tar})=\mathbb{P}(\mathbf{k}_1; \cdots; \mathbf{k}_T)$. In general, two or more blocking vectors may also be dependent as nearby targets can experience similar blocking. Thus, the joint distribution $\mathbb{P}(\mathbf{k_1}; \cdots; \mathbf{k}_T)$ captures correlated blocking in its entirety. Consequently, target-by-target localization is not optimal, in general. However, for ease of computation, we resort to such an approach in this paper, thereby implicitly assuming independent blocking vectors at distinct locations, i.e., $\mathbb{P}(\mathbf{k}|\Theta_{\rm tar}) \approx \displaystyle\prod\limits_t \mathbb{P}(\mathbf{k}_t)$. The generalization to joint-target localization will be described in a future work.

Among the $2^I$ possible values, $\mathbf{k}_t$ can only take on $(2^{M_{\rm TX}}-1)(2^{M_{\rm RX}}-1)+1$ physically realizable values, which can be expressed in the form $\mathbf{w}_t \bigotimes \mathbf{v}_t$ (e.g., $\mathbf{k}_t=[1~1~0~1~1~0~1~1~0]=[1~1~1]\bigotimes[1~1~0]$ for the TRP indexing notation in Table~\ref{tab:TRPeg}). These are referred to as \emph{consistent} blocking vectors while the remaining values are \emph{inconsistent} (e.g., $\mathbf{k}_t=[1~1~0~1~1~1~1~0~0]$). If $\mathbf{k}_t$ is inconsistent, then $\mathbb{P}(\mathbf{k}_t)=0$.

To characterize $\mathbb{P}(\hat{k}_{it}|k_{it})$, a distinction between two kinds of estimation errors needs to be made:
\begin{itemize}
 \item[a)] The DP corresponding to the $t$-th target at the $i$-th TRP may not detected if the noise pushes the range estimate far away from the true value. As a result, $\hat{k}_{it}=0$ when $k_{it}=1$. If the noise is independent and identically distributed (i.i.d) for all TRPs, we may assume that $\mathbb{P}(\hat{k}_{it}=0|k_{it}=1)=\rho_{01}$ ($\forall \hspace{2mm} i,t$), where $\rho_{01}$ is determined by the SNR (signal-to-noise ratio) and the ToA estimator.
 
 \item[b)] If the DP for the $t$-th target at the $i$-th TRP is blocked, but a noise peak or IP is mistaken for a DP because it has the right range, then $\hat{k}_{it}=1$ and $k_{it}=0$. $\mathbb{P}(\hat{k}_{it}=1|k_{it}=0)$ depends on the scatterer distribution and varies according to TX, RX and target locations. However, in the absence of IP statistics, we make the simplifying assumption that $\mathbb{P}(\hat{k}_{it}=1|k_{it}=0)=\rho_{10}$, for all $i,t$. The availability of empirical IP statistics would obviously improve localization performance.
\end{itemize}
Let $\hat{\mathbf{k}}_t = [\hat{k}_{1t}, \hspace{1mm} \cdots, \hspace{1mm}  \hat{k}_{It}]$ denote the estimated blocking vector at $(x_t,y_t)$. While $\hat{\mathbf{k}}_t$ can, in principle, take on all $2^I$ values, a false alarm is less likely if $\hat{\mathbf{k}}_t$ is a short Hamming distance away from a consistent vector having high probability. Let $\mathcal{K}$ denote the set of consistent blocking vectors. We restrict $\hat{\mathbf{k}}_t$ to be at most a unit Hamming distance from some element in $\mathcal{K}$. This assumption is reasonable when the number of scatterers is small and the SNR at all RXs is sufficiently high. Given $\hat{\mathbf{k}}_t$, let $\mathcal{K}_{t} \subseteq \mathcal{K}$ denote the set of consistent vectors that are at most a unit Hamming distance away from $\hat{\mathbf{k}}_t$. Then,
\begin{align}
 \mathbb{P}(\hat{\mathbf{k}}_t)&= \displaystyle\sum\limits_{\mathbf{k}_t \in \mathcal{K}_t} \mathbb{P}(\hat{\mathbf{k}}_t|\mathbf{k}_t)\mathbb{P}(\mathbf{k}_t) \notag\\
 \label{eq:hatk}
 &= \displaystyle\sum\limits_{\mathbf{k}_t \in \mathcal{K}_t} \left(\displaystyle\prod\limits_i \mathbb{P}(\hat{k}_{it}|k_{it})  \right) \mathbb{P}(\mathbf{k}_t) \\
 \label{eq:kerror}
 &\approx \begin{cases}
  \displaystyle\sum_{\mathbf{k}_t \in \mathcal{K}_{t}} \rho_{01}^{\eta_{01}} (1-\rho_{01})^{\eta_{11}} \rho_{10}^{\eta_{10}} (1-\rho_{10})^{\eta_{00}} \mathbb{P}(\mathbf{k}_t), \\
  \hspace{30mm} \mbox{if $\mathcal{K}_{t}$ is non-empty} \\
  0 , \hspace{26mm} \mbox{otherwise} \\
 \end{cases} \\
 \mbox{where,} ~ \eta_{01} &= |\{ i: \hat{k}_{it}=0; k_{it} = 1 \}| \notag\\
   \eta_{11} &= |\{ i: \hat{k}_{it}=1; k_{it} = 1 \}|\notag\\
 \eta_{10} &= |\{ i: \hat{k}_{it}=1; k_{it} = 0 \}| \notag\\
  \eta_{00} &= |\{ i: \hat{k}_{it}=0; k_{it} = 0 \}| \notag
\end{align}

Using (\ref{eq:hatk}) and assuming independent blocking vectors at distinct points (i.e., target-by-target detection), $P2$ can be reduced to the Bayesian MTL problem $P3$, given below:
\begin{align}
  \label{eq:Bayesobj_approx2}
&P3: \underset{T,\tilde{\mathbf{k}},\Theta_{\rm tar}}{\mbox{minimize}}~ \left[ \frac{1}{\sigma^2}\left(\displaystyle\sum\limits_{i,j,t} \tilde{k}_{ijt}(r_{ij} -r_i(x_t,y_t))^2 \right) \right. \notag\\ 
&\left. - \left(\displaystyle\sum\limits_{i,j,t} \tilde{k}_{ijt} \right)  \log\sqrt{2\pi}\sigma \right] - \displaystyle\sum\limits_t \log \mathbb{P}(\hat{\mathbf{k}}_t)  \\
& \mbox{subject to} ~ (\ref{eq:con1_approx1}), (\ref{eq:con2_approx1}) \notag
\end{align}
A \emph{matching} $q_t(\mathbf{r})=\{r_{ij} \in \mathbf{r} | \tilde{k}_{ijt}=1 \}$ is the set of DPs corresponding to the $t$-th target. Given $q_t(\mathbf{r})$ and a point $(x_t,y_t)$, the term in square parentheses in (\ref{eq:Bayesobj_approx2}) determines if the ellipses corresponding to the MPCs in $q_t(\mathbf{r})$ pass through $(x_t,y_t)$ or not. The other term in (\ref{eq:Bayesobj_approx2}) plays the role of a prior by determining the probability of the blocking vector, $\hat{\mathbf{k}}_t$, obtained from $q_t(\mathbf{r})$, at $(x_t,y_t)$. The objective in (\ref{eq:Bayesobj_approx2}) is minimized only when both these quantities are small. To solve $P3$, a mechanism for detecting DPs is required. Since the IP distribution is unknown, none of the conventional tools such as Bayesian, minimax or Neyman-Pearson hypothesis testing can be used for this purpose. In the next section, we describe our DP detection technique and propose a polynomial-time algorithm to solve $P3$.

\section{MTL Algorithm using Blocking Statistics}
\label{sec:algo}
In this section, we define a likelihood function for identifying DPs that enable us to obtain the matchings required for solving $P3$ in a tractable manner.

The number of matchings possible for $T$ targets, $M$ scatterers and $I$ TRPs is $({I\choose 3} N^3 + {I\choose 4} N^4 \cdots + {I\choose I} N^I)^T$, where $N=(2M+1)T$ is an upper bound on the number of MPCs extracted at each TRP, ignoring noise peaks. The computational complexity of a brute-force search over all possible matchings for solving $P3$ is $O(N^{IT})$, which is intractable for a large number of TRPs and/or targets. To obtain accurate matchings in a tractable manner, we employ an iterative approach. Consider, without loss of generality, a matching $q_t^{(i-1)}(\mathbf{r})$ for the $t$-th target consisting of MPCs from the first $i-1$ TRPs ($3 \leq i \leq I$). The size of $q_t^{(i-1)}(\mathbf{r})$ is at most $i-1$. Let $(\hat{x}_t^{(i-1)},\hat{y}_t^{(i-1)})$ denote the estimate of the target location obtained from $q_t^{(i-1)}(\mathbf{r})$ (e.g., using the two-step estimation method \cite{Shen_et_al_2012}). For an MPC $r_{ij_i}$ from the $i$-th TRP, let $q_{t,\rm temp}^{(i)}(\mathbf{r}) = q_{t}^{(i-1)}(\mathbf{r}) \cup r_{ij_i}$ and let $\hat{\mathbf{k}}_t^{(i)}$ denote the $i$-length partial blocking vector at $(\hat{x}_t^{(i-1)},\hat{y}_t^{(i-1)})$, obtained from $q_{t,\rm temp}^{(i)}(\mathbf{r})$. If $q_{t,\rm temp}^{(i)}(\mathbf{r})$ consists entirely of DPs from $(x_t,y_t)$, then (i) the ellipses corresponding to its constituent MPCs should pass \emph{close} to $(x_t,y_t)$, and (ii) the blocking vector $\hat{\mathbf{k}}_t^{(i)}$ should have high probability. This motivates the definition of a blocking-aware vector likelihood function, $\mathbf{L_B}(q_{t,\rm temp}^{(i)}(\mathbf{r}))$, defined as follows:
\begin{align}
\label{eq:newLfnc}
 \mathbf{L_B}(q_{t,\rm temp}^{(i)}(\mathbf{r}))&= \left(\left|\frac{L_{\rm E}(q_{t,\rm temp}^{(i)}(\mathbf{r}))}{\sigma(q_{t,\rm temp}^{(i)}(\mathbf{r}))}\right|, -\log \mathbb{P} (\hat{\mathbf{k}}^{(i)}_t)\right) \\
  \label{eq:oldLfnc}
 \mbox{where,}~ L_{\rm E}(q_{t,\rm temp}^{(i)}(\mathbf{r}))&= r_{ij_i}-r_i(\hat{x}_t^{(i-1)},\hat{y}_t^{(i-1)})
\end{align} 
and $\sigma(q_{t,\rm temp}^{(i)}(\mathbf{r}))$ is the standard deviation of $L_{\rm E}(q_{t,\rm temp}^{(i)}(\mathbf{r}))$. 

If the ellipses corresponding to the MPCs in $q_{t,\rm temp}^{(i)}(\mathbf{r})$ pass through the vicinity of $(x_t,y_t)$, then $L_{\rm E}(q_{t,\rm temp}^{(i)}(\mathbf{r}))$ should be very small in magnitude. Under this condition, it can be shown by a Taylor's series approximation that $L_{\rm E}(q_{t,\rm temp}^{(i)}(\mathbf{r}))$ is a zero-mean Gaussian random variable \cite{Shen_and_Molisch_2013_2}. Hence, if $|L_{\rm E}(q_{t,\rm temp}^{(i)}(\mathbf{r}))/\sigma(q_{t,\rm temp}^{(i)}(\mathbf{r}))| \leq \delta$, where $\delta$ is an \emph{ellipse intersection threshold}, then we conclude that $r_{ij_i}$ passes through $(\hat{x}_t^{(i-1)},\hat{y}_t^{(i-1)})$.

If the above ellipse intersection condition is satisfied, then the term $-\log \mathbb{P} (\hat{\mathbf{k}}^{(i)}_t)$, which denotes the \emph{blocking likelihood} of $q_{t,\rm temp}^{(i)}(\mathbf{r})$ at $(\hat{x}_t^{(i-1)},\hat{y}_t^{(i-1)})$, needs to be small as well. The following cases are of interest:
\begin{itemize}
 \item[1.] If $\hat{\mathbf{k}}_t^{(i)}$ is consistent and $-\log \mathbb{P} (\hat{\mathbf{k}}^{(i)}_t) \leq \mu$, where $\mu (>0)$ is a \emph{blocking threshold}, then we define $q_t^{(i)}(\mathbf{r})=q_{t,\rm temp}^{(i)}(\mathbf{r})$ and compute a refined target location estimate $(\hat{x}_t^{(i)},\hat{y}_t^{(i)})$ from $q_t^{(i)}(\mathbf{r})$.
 \item[2.] If $\hat{\mathbf{k}}^{(i)}_t$ is inconsistent, then let $\mathcal{K}_t^{(i)}$ denote the set of consistent $i$-length partial blocking vectors that are at most a unit Hamming distance away from $\hat{\mathbf{k}}^{(i)}_t$. The following cases are of interest then:
 \begin{itemize}
 \item[(a)] If $\mathcal{K}_t^{(i)}$ is empty, then $\mathbb{P}(\hat{\mathbf{k}}^{(i)}_t)=0$ (from (\ref{eq:hatk}) and (\ref{eq:kerror}), which hold for partial blocking vectors as well) and $-\log \mathbb{P}(\hat{\mathbf{k}}^{(i)}_t) = \infty$. Hence, we conclude that a target is not present at the estimated location.
 \item[(b)] If $\mathcal{K}_t^{(i)}$ is not empty, then each element of $\mathcal{K}_t^{(i)}$ is a feasible ground truth. In particular, an element in $\mathcal{K}_t^{(i)}$ whose Hamming weight is lower than that of $\hat{\mathbf{k}}_t^{(i)}$ represents a ground truth where exactly one MPC in $q_{t,\rm temp}^{(i)}(\mathbf{r})$ is not a DP. For each such element, a new matching can be derived by removing the corresponding \emph{non-DP} from $q_{t,\rm temp}^{(i)}(\mathbf{r})$ and evaluated a new blocking likelihood. On the other hand, an element of $\mathcal{K}_t^{(i)}$ with a higher Hamming weight compared to $\hat{\mathbf{k}}_t^{(i)}$ represents a ground truth where one DP is absent from $q_{t,\rm temp}^{(i)}(\mathbf{r})$ due to noise. Unlike the previous case, no modification of the matching is possible and the blocking likelihood of $\hat{\mathbf{k}}_t^{(i)}$ is computed according to (\ref{eq:hatk})-(\ref{eq:kerror}). In this manner, it is possible that multiple matchings may exist for a single potential target location, each corresponding to a different ground truth. All the matchings whose blocking likelihood satisfies the threshold $\mu$ are retained, since it is premature to determine the most likely ground truth until all TRPs are considered. After the $I$-th TRP has been processed, if multiple matchings still exist for the $t$-th target, then the one that minimizes the objective function in (\ref{eq:Bayesobj_approx2}) is declared the true matching and the corresponding $(\hat{x}_t^{(I)},\hat{y}_t^{(I)})$ is the location estimate for the $t$-th target.
 \end{itemize}
\end{itemize}
Otherwise, if no MPC from the $i$-th TRP satisfies the ellipse intersection condition (i.e., 
$|L_{\rm E}(q_{t,\rm temp}^{(i)}(\mathbf{r}))/\sigma(q_{t,\rm temp}^{(i)}(\mathbf{r}))|$ $>$ $\delta$, for all $r_{ij_i}$), then $q_t^{(i)}(\mathbf{r}) = q_t^{(i-1)}(\mathbf{r})$ and $(\hat{x}_t^{(i)},\hat{y}_t^{(i)})=(\hat{x}_t^{(i-1)},\hat{y}_t^{(i-1)})$. For the resulting $\hat{\mathbf{k}}_t^{(i)}$, inconsistencies are handled as stated above in point 2. If $-\log \mathbb{P}(\hat{\mathbf{k}}_t^{(i)})> \mu$, then we conclude that a target is not present at the estimated location.

This motivates an algorithmic approach that is divided into stages, indexed by $i$. In general, let $(z_1,z_2,...,z_I)$, a permutation of $(1, 2, \cdots, I)$, be the order in which TRPs are processed. At the beginning of the $i$-th stage $(3\leq i \leq I)$, each $q_t^{(i-1)}(\mathbf{r})$ has at most $i-1$ entries. During the $i$-th stage, all the DPs among the MPCs of the $z_i$-th TRP are identified to obtain a set of matchings $\{q_t^{(i)}(\mathbf{r})\}$ for each target $t$. A matching is consistent (inconsistent) if the corresponding blocking vector is consistent (inconsistent). By construction, the only inconsistent matchings are due to missing DPs (see bullet point 2(b) in previous paragraph). A finite value of $\mu$ ensures that an inconsistent $\hat{\mathbf{k}}^{(i)}_t$ is always a unit Hamming distance away from consistency, due to (\ref{eq:kerror}). Since the blocking likelihood $-\log \mathbb{P}(\hat{\mathbf{k}}_t^{(i)})$ is non-decreasing in $i$, a matching and its corresponding target location can be removed from consideration if at any stage its blocking likelihood exceeds the blocking threshold, $\mu$.

Let $P(a,b,j_a,j_b)$ denote the points of intersection of the ellipses corresponding to the $j_a$-th MPC of the $a$-th TRP and the $j_b$-th MPC of the $b$-th TRP. For the initial set of matchings (i.e., $i=3$), $P(z_1,z_2,j_{z_1},j_{z_2})$ is computed for all $j_{z_1}, j_{z_2} (1 \leq j_{z_1} \leq N_{z_1}, 1 \leq j_{z_2} \leq N_{z_2})$. There can be at most four points in any $P(z_1,z_2,j_{z_1},j_{z_2})$ and each such point is an ML estimate of the target location for the matching $q_t^{(2)}(\mathbf{r})=\{r_{z_1,j_{z_1}}, r_{z_2,j_{z_2}}\}$. Hence, the target location estimate $(\hat{x}_t^{(2)},\hat{y}_t^{(2)})$ need not be unique. Furthermore, in the $i$-th stage ($i\geq 4$) we also compute $P(z_u,z_i,j_{z_u},j_{z_i})$ for all $j_{z_u},j_{z_i} (\forall u<i)$ to identify previously blocked targets.

In summary, any intersection of two ellipses is a potential target location to begin with. At each such location, the likelihood of a target being present is updated depending on the number of other ellipses passing around its vicinity. Unlikely target locations, corresponding to matchings whose likelihood (given by $\mathbf{L}_{\mathbf{B}}(.)$) does not satisfy the thresholds $\delta$ and $\mu$, are eliminated at each stage. The number of targets that remain at the end is the estimate of $T$. Algorithm \ref{algo} lists the pseudocode of the Bayesian MTL algorithm.

				        		
\subsection{Complexity of Bayesian MTL Algorithm}
Let $\hat{T}(i)$ denote the number of targets identified at the end of stage $i$. The following relation holds,
\begin{align}
 \label{eq:T_est}
 \hat{T}(i) &\leq \hat{T}(i-1) + {i-1 \choose 2}N^3, \hspace{2mm} (i=4, \cdots, I) \\
 \mbox{and,} \hspace{3mm} \hat{T}(3) &\leq N^3
\end{align}
At the end of $i-1$-th stage, each target can have at most $(i-1)$ matchings. Hence, $O(i\hat{T}(i-1)N)$ likelihood computations are carried out in the $i$-th stage due to existing targets. The second term in (\ref{eq:T_est}) is an upper bound on the number of new targets that can be identified in the $i-th$ stage and the number of likelihood computations due to these is $O(i^2 N^3)$. At each stage, the number of potential targets increases at most polynomially in $N$ and $I$ (\ref{eq:T_est}). Hence, the number of likelihood computations is also polynomial in $N$ and $I$. The reduction in complexity occurs because target locations are determined by `grouping' pair-wise ellipse intersections that are close together. Since there are only $O(I^2N^2)$ ellipse intersections to begin with, it is intuitive that the proposed algorithm terminates in polynomial-time.

\subsection{Limitations of Bayesian MTL Algorithm}
The Bayesian MTL algorithm assumes complete knowledge of the distribution of $\mathbf{k}_t$ at all locations $(x_t,y_t)$. This would have to be obtained either from very detailed theoretical models or exhaustive measurements, neither of which might be feasible in practice. A sub-optimal, but more practical, alternative could involve the use of second-order statistics of $\mathbf{k}_t$. In particular, the Mahalanobis distance, defined as $\sqrt{(\hat{\mathbf{k}}_t-\mathbf{m}_t)^T\mathbf{C}_t^{-1}(\hat{\mathbf{k}}_t-\mathbf{m}_t)}$, where $\mathbf{m}_t$ and $\mathbf{C}_t$ respectively denote the mean vector and covariance matrix of $\mathbf{k}_t$ and $(.)^T$ and $(.)^{-1}$ denote the matrix transpose and inverse operations, respectively, can be compared to a threshold $\mu_2$ as the basis for a blocking likelihood decision. Even in this simplified case, one still needs the mean blocking vector and the covariance matrix at each point. In practice, these can be measured at only at a fixed set of grid points. Hence, the accuracy of the algorithm would depend on the grid resolution of the measured data.
\begin{algorithm}[t]
 \caption{Bayesian MTL algorithm}
 \begin{algorithmic}
  \State Obtain the TRP processing order $(z_1,z_2,\cdots,z_n)$ \cite{Shen_and_Molisch_2013_2}
  \State $t=0$
  \Comment {{\color{red} (Initial set of matchings)}}
  \For {each $j_{z_1}, j_{z_2}$}
  	\For {each ellipse intersection $(x,y)$ corresponding to $r_{z_1,j_{z_1}}$ and $r_{z_2,j_{z_2}}$}
  		\If {$\mathbf{L_B}(\{r_{z_1,j_{z_1}},r_{z_2,j_{z_2}}\}) \leq (\delta,\mu)$}
  		  \State $t=t+1$ 
  		  \State $q_t^{(2)}(\mathbf{r})=\{r_{z_1,j_{z_1}},r_{z_2,j_{z_2}}\}$
  		  \State $(\hat{x}_t^{(2)},\hat{y}_t^{(2)})=(x,y)$
  		\EndIf
  	\EndFor
  \EndFor	
  \State $\hat{T}(2)=t$	\Comment {{\color{red} $\hat{T}(i)$ denotes the number of estimates at the end of the $i$-th stage}}  
  \For {$i=3$ to $I$}
	   \For {$t=1$ to $\hat{T}(i-1)$}
	   		\Comment {{\color{red} (Updating existing matchings)}}
	   		\If {$\exists$ any $r_{z_i,j_{z_i}}$ such that $L_E(q_{t}^{(i-1)}(\mathbf{r})\cup r_{z_i,j_{z_i}})\leq\delta$}
	   			\State $q_{t,\rm temp}^{(i)}(\mathbf{r})=q_t^{(i-1)}(\mathbf{r}) \cup r_{z_i,j_{z_i}}$
	   		\Else
	   			\State $q_{t,\rm temp}^{(i)}(\mathbf{r})=q_t^{(i-1)}(\mathbf{r})$
	   		\EndIf		
	   		\State Derive $\mathcal{K}_t^{(i)}$ from $q_{t,\rm temp}^{(i)}(\mathbf{r})$
	 		\For {each $\hat{\mathbf{k}}_t^{(i)} \in \mathcal{K}_t^{(i)}$}
 		    	\If {$-\log \hat{\mathbf{k}}_t^{(i)} \leq \mu$} 
	    			\State Derive $q_t^{(i)}(\mathbf{r})$ from $q_{t,\rm temp}^{(i)}(\mathbf{r})$ according to $\hat{\mathbf{k}}_t^{(i)}$
	    		\EndIf
	    	\EndFor
   		\EndFor 
   		\State Update $\hat{T}(i)$ and set $t=\hat{T}(i)$

	    \For {each $j_{z_i}, j_{z_u}$ ($u=1,\cdots,i-1$)}
	   		\Comment {{\color{red} (New targets, previously unidentified due to blocking)}}
 		  	\For {each ellipse intersection $(x,y)$ corresponding to $r_{z_i,j_{z_i}}$ and $r_{z_u,j_{z_u}}$}
 		  	
		  		\If {$\mathbf{L_B}(\{r_{z_i,j_{z_i}},r_{z_u,j_{z_u}}\}) \leq (\delta,\mu)$}
		  		  \State $t=t+1$ 
		  		  \State $q_t^{(i)}(\mathbf{r})=\{r_{z_i,j_{z_i}},r_{z_u,j_{z_u}}\}$
		  		  \State $(\hat{x}_t^{(i)},\hat{y}_t^{(i)})=(x,y)$
		  		\EndIf
  			\EndFor
  		\EndFor	
  		\State $\hat{T}(i)=t$
  \EndFor
 \end{algorithmic}
 \label{algo} 
\end{algorithm}

\section{Simulation and Experimental Results}
\label{sec:simres}
In this section, we present our simulation and experimental results for the Bayesian MTL algorithm introduced in Section~\ref{sec:algo}. In Section~\ref{sec:priorart}, the algorithm is validated by reproducing the results described in prior art for independent blocking, which is a special instance of $P3$. The importance of considering correlated blocking and the accuracy of the matchings obtained by the Bayesian MTL algorithm are discussed in Sections~\ref{sec:mainsim} and \ref{sec:genie}, respectively. Finally, experimental results which provide insights into the impact of non-point targets and imperfect background subtraction are presented in Section~\ref{sec:exp}.

Unless otherwise mentioned, we use the following settings for our simulation results: $G = [-10{\rm m}, 10{\rm m}] \times [-10{\rm m}, 10{\rm m}]$ is the region of interest. Scatterers are modelled as balls of diameter $L$; obviously, the blocking correlation increases with $L$. The standard deviation of the ranging error, $\sigma$ is assumed to be $0.01 {\rm m}$. Two or more MPCs that are within a distance of $2\sigma$ apart are considered to be unresolvable; in that case, the earliest arriving peak is retained and the other peaks are discarded. For a given $\delta$, $\rho_{01}=\rho_{10}=2Q(\delta)$ was assumed, where $Q(x)=\displaystyle\int\limits_x^{\infty} e^{-x^2/2}/\sqrt{2\pi} dx$. A target is considered to be missed if there is no location estimate lying within a radius of $3\sigma$ from the actual coordinates. Similarly, a false alarm is declared whenever there is no target within a radius of $3\sigma$ from an estimated target location. For a given network realization, let $\hat{T}_D$ and $\hat{T}_F$ denote the number of detections and false alarms, respectively. Then, the detection and false alarm probabilities, denoted by $P_D$ and $P_F$, respectively, are calculated as follows,
\begin{align}
 P_D &= \mathbb{E}[\hat{T}_D/T] \\
 P_F &= \mathbb{E}[\hat{T}_F/(\hat{T}_D+\hat{T}_F)]
\end{align}
where the expectation is over the ensemble of network realizations

\subsection{Comparison to Prior Art}
\label{sec:priorart}
In \cite{Shen_and_Molisch_2013_2}, the probability of \emph{any} DP being blocked was assumed to be constant throughout $G$ and independent of other blocking probabilities. Target detection was achieved if there existed a matching of size at least $I-\Phi$, where $\Phi$ denotes the maximum number of \emph{undetected} DPs permitted, regardless of consistency. We now proceed to demonstrate how this criterion is a special case of the Bayesian MTL Algorithm, obtained by assuming independent blocking with constant blocking probabilities (henceforth referred to as the i.c.b assumption). Let $p_{\rm los}$ denote the probability that LoS exists between any two points in $G$. The probability that a DP is blocked is then given by $p_{\rm b}=1-p_{\rm los}^2$. Taking into account both blockage and missed detection by noise, the probability that a DP is undetected (denoted by $p_{\rm dp}$) is given by
\begin{align}
 \label{eq:pdp}
 p_{\rm dp} = (1-p_{\rm b}).2Q(\delta)+p_{\rm b}
\end{align}
The blocking likelihood of a matching with $\Phi$ undetected DPs equals $-\log ((1-p_{\rm dp})^{I-\Phi} p_{\rm dp}^{\Phi})$. If $p_{\rm dp} < 1/2$, then the blocking likelihood monotonically increases with $\Phi$. Hence, for a given $\Phi$, the corresponding blocking threshold, $\mu(\Phi)$, can be set as follows
\begin{align}
\label{eq:mu(phi)}
 \mu (\Phi)= -\log ((1-p_{\rm dp})^{I-\Phi} p_{\rm dp}^{\Phi})
\end{align}
which ensures that the detected targets have matchings of size at least $I-\Phi$.

\begin{figure}
 \includegraphics[scale=0.55]{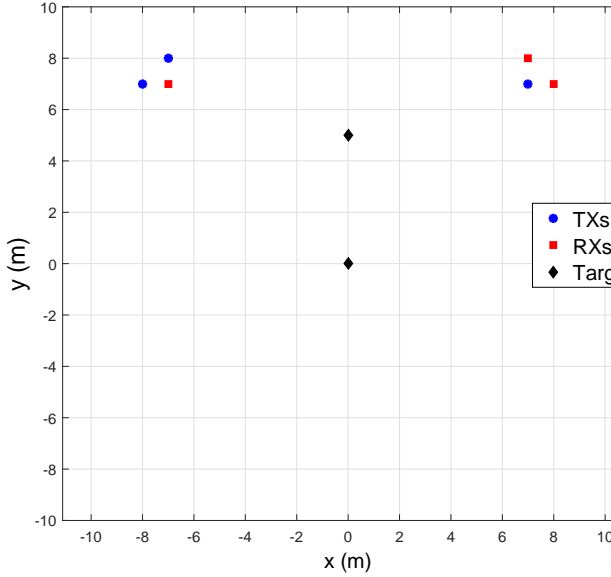}
 \caption{A network consisting of 3 TXs at $(-8{\rm m},7{\rm m})$, $(-7{\rm m},8{\rm m})$ and $(7{\rm m},7{\rm m})$, 3 RXs at $(-7{\rm m},7{\rm m})$, $(8{\rm m},7{\rm m})$ and $(7{\rm m},8{\rm m})$ (i.e., $I=9$ TRPs) and 2 targets at $(0{\rm m},0{\rm m})$ and $(0{\rm m},5{\rm m})$. The TX and RX locations are such that the LoS blocking probabilities are independent only if $L$ is very small. For $L=0.001{\rm m}$, the independent blocking assumption holds.}
 \label{fig:contrived_network}
\end{figure}

\begin{figure}
 \includegraphics[scale=0.5]{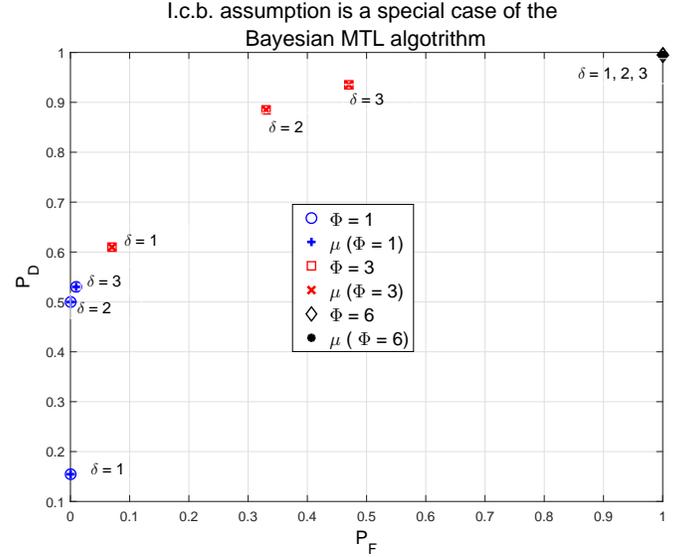}
 \caption{Region of Convergence (ROC) curves plotting $P_D$ versus $P_F$ for the i.c.b condition. The prior art in \cite{Shen_and_Molisch_2013_2} is a special case of the Bayesian MTL algorithm}
 \label{fig:priorart}
\end{figure}

To validate the Bayesian MTL algorithm, we compared it with the prior art proposed in \cite{Shen_and_Molisch_2013_2}, under the i.c.b assumption. The comparison was done on the network shown in Fig. \ref{fig:contrived_network}. To model the i.c.b condition, the values for $L$ and $p_{\rm los}$ were chosen to be $0.001{\rm m}$ and 0.9, respectively. With probability $1-p_{\rm los}$, a scatterer was placed independently and uniformly along each line segment between a node (TX/RX) and a target.
The two algorithms were evaluated over 100 realizations for three values of $\delta(= 1, 2 \mbox{ and } 3)$  and $\Phi$ $(=1, 3$ and $6)$. For each value of $\Phi$, the threshold $\mu(\Phi)$ for the Bayesian MTL algorithm was chosen according to (\ref{eq:mu(phi)}). The region of convergence (ROC) curves, plotting $P_D$ versus $P_F$, for both algorithms are shown in Fig. \ref{fig:priorart}. As expected, they yield identical missed-detection and false alarm rates.

Increasing $\delta$ loosens the compactness constraint on the ellipse intersections around a potential target location, while increasing $\mu$ relaxes the constraint on the probability of a blocking vector/matching. Hence, both $P_F$ and $P_D$ are non-decreasing in $\delta$ and $\mu$, as seen in Fig. \ref{fig:priorart}. In the special case where only three ellipse intersections are sufficient to declare the presence of a target $(\Phi=6)$, the false alarm rates are very high. This is in agreement with the results reported in \cite{Aditya_and_Molisch_2015}.

\subsection{Effect of Correlated Blocking}
\label{sec:mainsim}
\begin{figure}
 \includegraphics[scale=0.6]{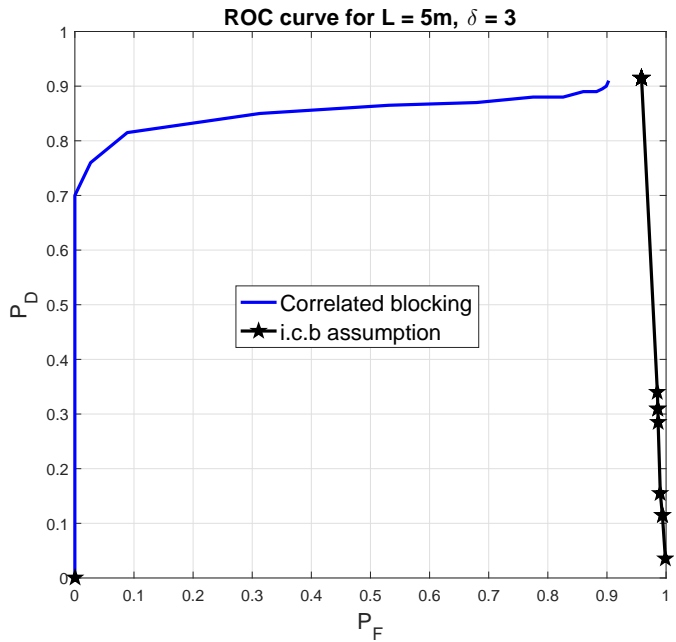}
 \caption{Ignoring correlated blocking can result in false alarms being more likely to occur than detections.}
 \label{fig:ROC_justify}
\end{figure}

To highlight the effect of correlated blocking, the value of $L$ was increased to $5{\rm m}$ and the scatterer centers were distributed according to a homogeneous Poisson point process (PPP) of intensity $\lambda= 0.0075{\rm m}^{-2}$, which amounts to three scatterers in $G$, on average, per realization. The blocking distribution for the PPP scatterer model is derived in Appendix \ref{app:blockmodel}. A total of 100 network realizations were considered, with $M_{\rm TX}=M_{\rm RX}=3$ and $T=2$, which corresponds to $N=T(2M+1)=2(2\times 3+1)=14$ MPCs per TRP, on average. Let $S_{\rm sc}\subseteq G$ denote the region occupied by scatterers in a given realization. The TX, RX and target locations were uniformly and independently distributed over the region $G\setminus S_{\rm sc}$, where `$\setminus$' denotes the set difference operator. Under the i.c.b assumption for the above settings, $p_{\rm los}=\exp(-\lambda L d_{\rm avg})$, where $d_{\rm avg}=10.1133 {\rm m}$ is the average distance between a target and a node. Hence, from (\ref{eq:pdp}), $p_{\rm dp}=0.5329 > 1/2$ for $\delta=3$. The distribution of the average number of DPs at a point is tabulated in Table \ref{tab:sumk_pmf} for both the true blocking distribution and the i.c.b assumption. As per the true blocking distribution, a target has LoS to all TXs and RXs (i.e., 9 DPs) over $66\%$ of the time and the probability that a target has only 3 DPs is a little over $1\%$. As a result, a matching of size 3 is more likely to be a false alarm. However, since $p_{\rm dp} > 1/2$, a matching of size 3 is more probable than a matching of size 9 (which occurs with less than $1\%$ probability) under the i.c.b assumption. As a result, false alarms are identified first, followed by detections, as the value of $\mu$ increases. This is reflected in the ROC curves plotted in Fig. \ref{fig:ROC_justify}, where the i.c.b assumption gives rise to very high false alarm rates.

\begin{table*}
 \centering
 \begin{tabular}{|c|c|c|c|c|c|c|c|c|}
  \hline
   Blocking Distribution & $< 3$ & 3 & 4 & 5 & 6 & 7 & 8 & 9 \\
  \hline
   True & 0.0700 & 0.0150 & 0.0750 & 0 & 0.1750 & 0 & 0 & 0.6650 \\
   i.c.b assumption & 0.3367 & 0.1961 & 0.2578 & 0.2259 & 0.1320 & 0.0496 & 0.0109 & 0.0011 \\
  \hline
 \end{tabular}
 \caption{Distribution of the average number of DPs at a point for $L=5{\rm m}$ and $\lambda=0.0075{\rm m^{-2}}$.}
 \label{tab:sumk_pmf}
\end{table*}

\subsection{Comparison with genie-aided method}
\label{sec:genie}
\begin{figure}
 \includegraphics[scale=0.6]{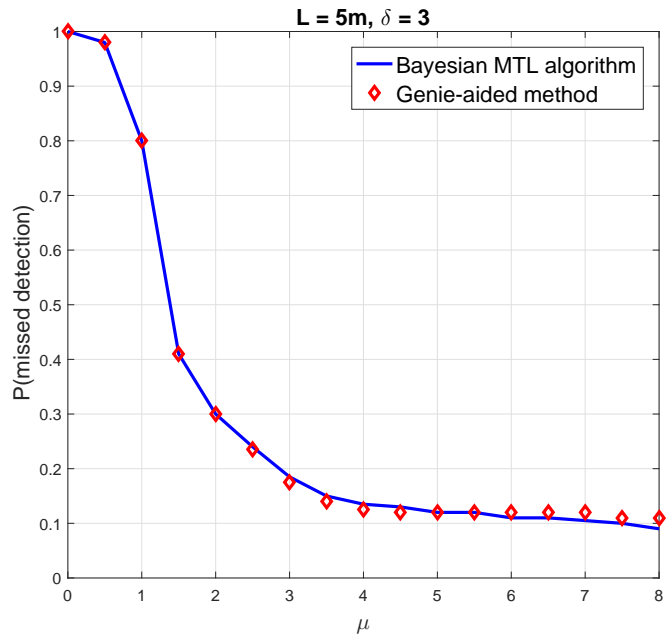}
 \caption{Comparison with genie-aided method.}
 \label{fig:genie_dia5}
\end{figure}

In many radar applications, a missed-detection is more costly than a false alarm. As a benchmark, the missed-detection probability of the Bayesian MTL algorithm is compared with that of a genie-aided method, which involves running the Bayesian MTL Algorithm on the true target matchings, in Fig. \ref{fig:genie_dia5}. It can be seen that the proposed algorithm performs as well as the genie-aided method.

\subsection{Experimental Results}
\label{sec:exp}
\begin{figure}
 \centering
 \begin{subfigure}{0.5\textwidth}
  \centering
 \includegraphics[scale=0.45]{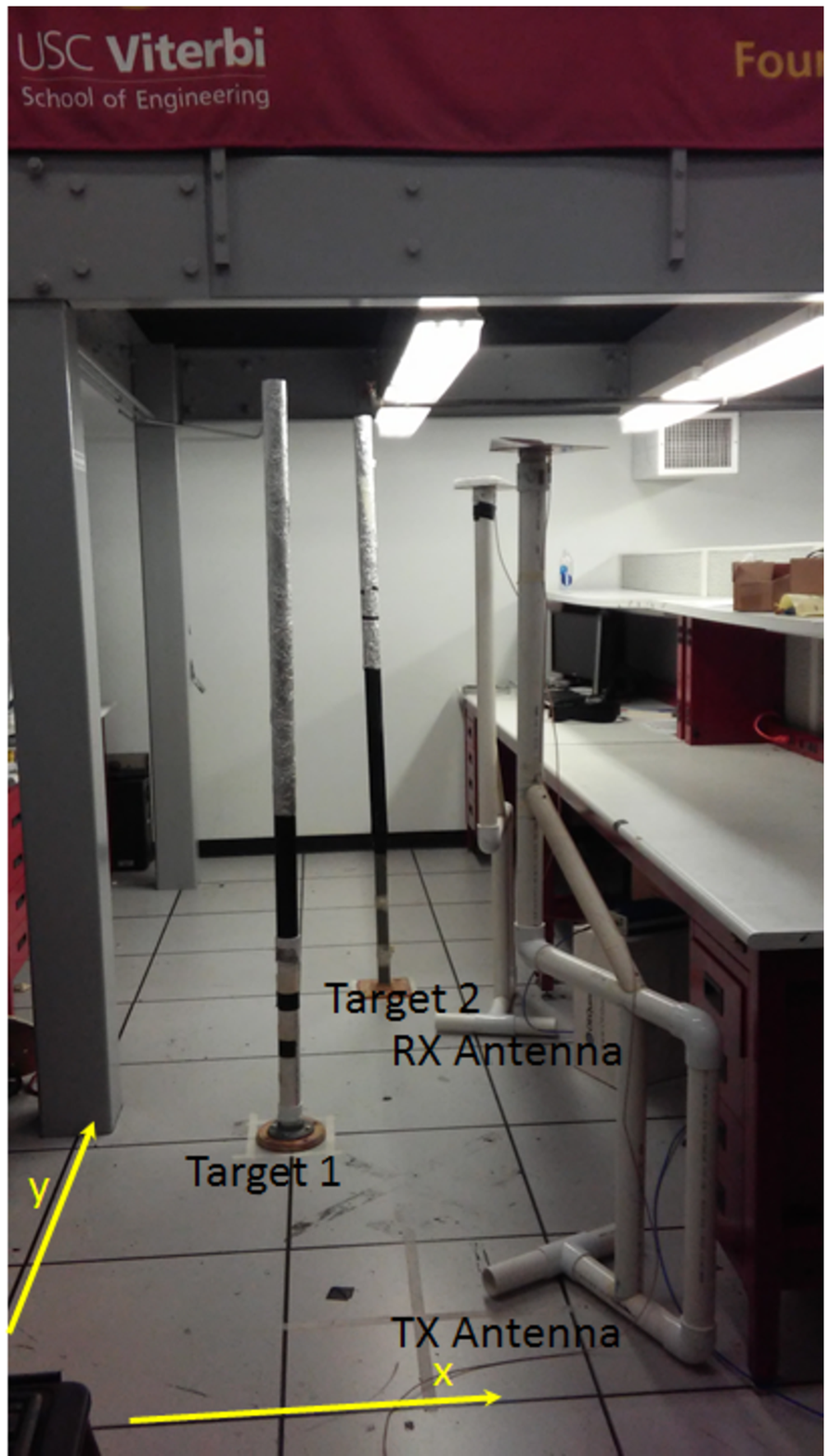} 
 \caption{The cluttered indoor measurement environment.}
 \label{fig:pic_env2}
\end{subfigure}
\\
\vspace{5mm}
\begin{subfigure}{0.5\textwidth}
 \includegraphics[scale=0.3]{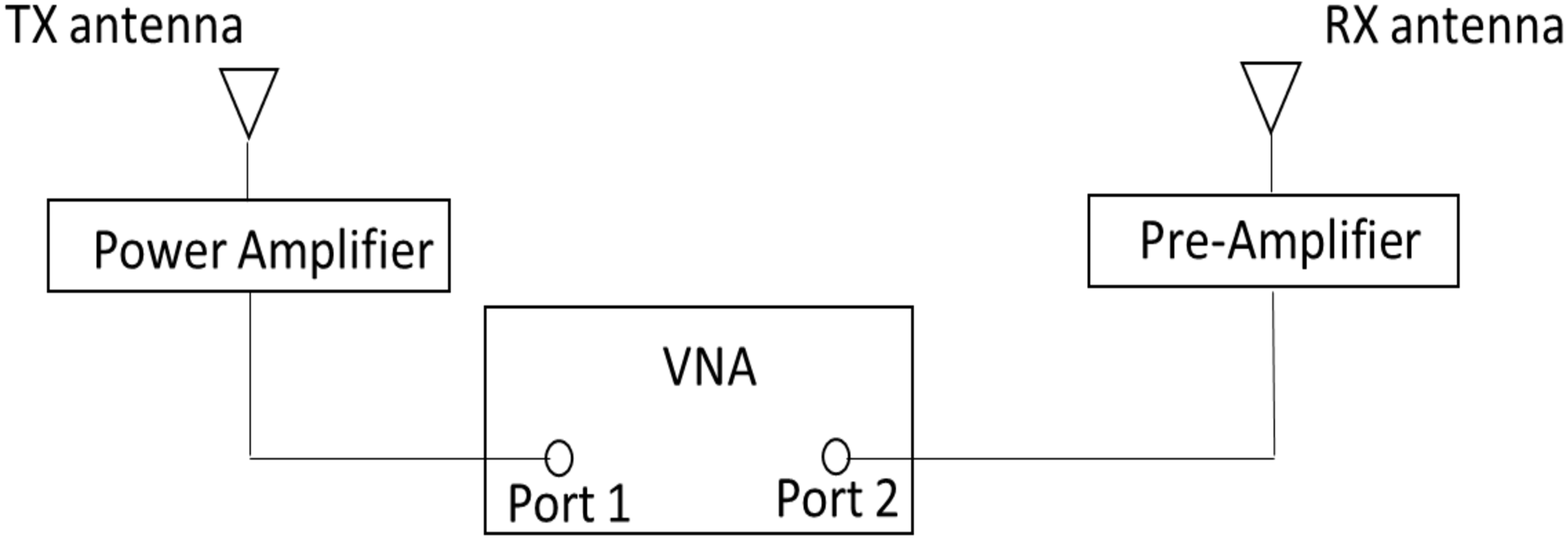} 
 \caption{Measurement setup using VNA}
 \label{fig:setup}
\end{subfigure}
\caption{The experimental setup.}
\end{figure}

We now present some experimental results that further validate the performance of the Bayesian MTL algorithm. We chose a portion of UltraLab at USC, a cluttered indoor environment, for our measurements. The floor was paved with square tiles of side $0.61\rm m$ which provided a natural Cartesian coordinate system, as shown in Fig.~\ref{fig:pic_env2}. The measurement setup is shown in Fig.~\ref{fig:setup}. For the $i$-th TRP, the frequency response of the ultrawideband (UWB) channel over 6-8 {\rm GHz} was measured twice at 1601 frequency points - once without the targets (i.e., the background measurement, denoted by $H^{\rm back}_i(f)$) and then with the targets present (denoted by $H^{\rm tar}_i(f)$) - using a pair of horn antennas with beamwidth $60^o$, connected to a vector network analyzer (VNA). This corresponds to $\sigma=0.15{\rm m}$. Horn antennas were preferred over omnidirectional antennas to restrict the background clutter to a narrow sector. The antennas were maintained at the same height from the ground in order to create a 2D localization scenario, and were oriented to face the targets. Two identical, foil-wrapped cylindrical poles were chosen as the targets. Although the height of the cylinders exceeded that of the TX and RX antennas, the portion of the cylinder that was in the plane of the antennas was wrapped in foil to maintain the 2D nature of the problem. 

\begin{figure}
 \includegraphics[scale=0.6]{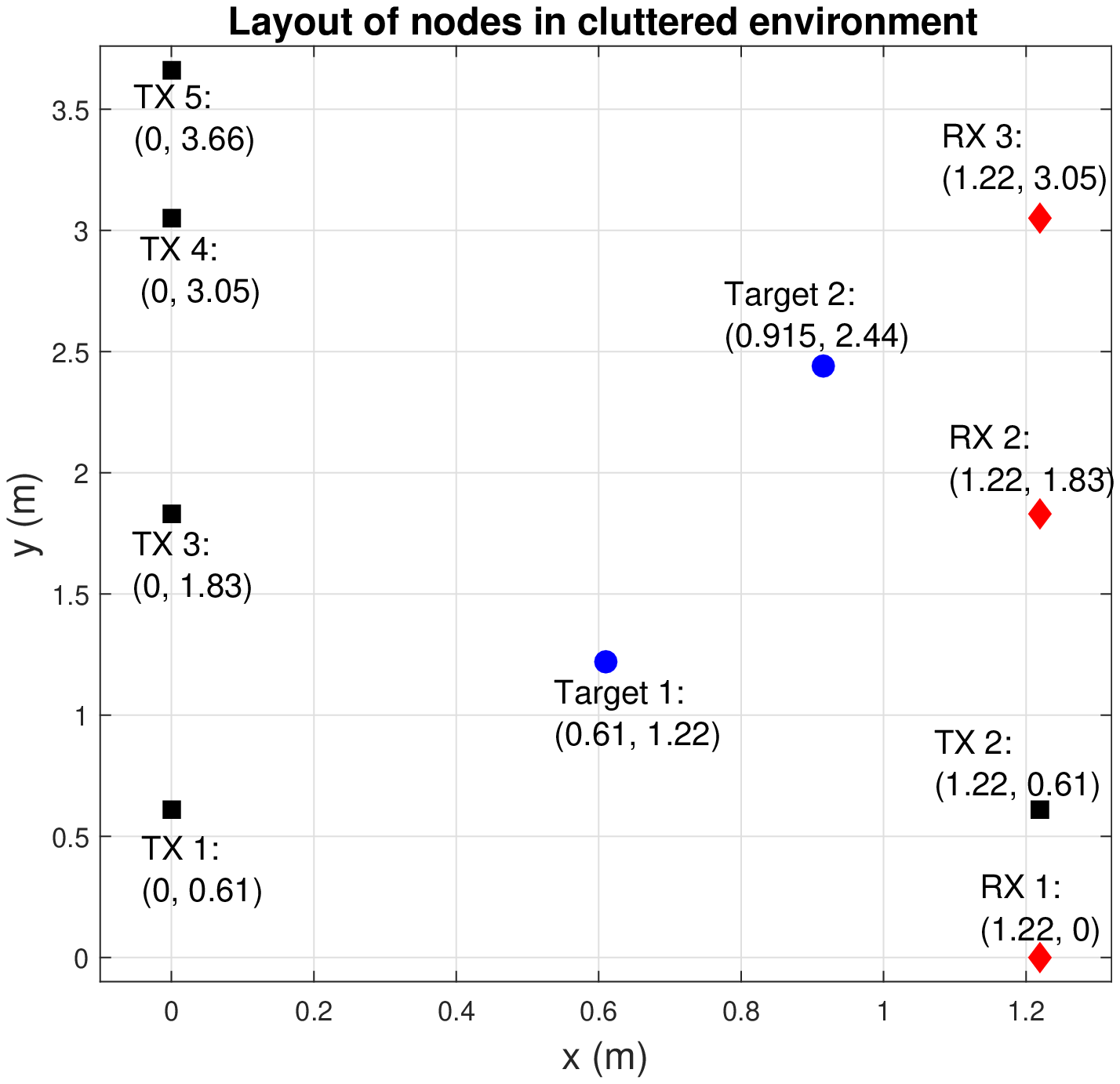} 
 \caption{Layout of TXs, RXs and targets in the cluttered environment of Fig.~\ref{fig:pic_env2}.}
 \label{fig:lay_env2}
\end{figure}

Let $h_i(t)$ denote the channel impulse response for the $i$-th TRP due to the targets alone (i.e., after background subtraction). Then, $h_i(t)$ is given by the following expression,
\begin{align}
 h_i(t)={\rm IFFT}(H^{\rm tar}_i(f) - H_i^{\rm back}(f))
\end{align}
The noise floor corresponding to the $i$-th TRP was determined by computing the average power in the last 100 samples of $h_i(t)$. These delay bins correspond to a signal run length in excess of $200\rm m$, which is well in excess of the ranges encountered in our measurement scenario (less than $10\rm m$). Hence, it is reasonable to assume that the energy in these delay bins is due to thermal noise alone. After determining the noise power, MPCs were extracted from $h_i(t)$ whenever the SNR was greater than $10{\rm dB}$. A distributed virtual MIMO radar was implemented by moving the TX and RX antennas to different locations, as shown in  Fig.~\ref{fig:lay_env2}. Six TRPs were considered, which are indexed in Table~\ref{tab:lookup_scenario2}. LoS was present between all TXs, RXs and targets (i.e., $\mathbb{P}(\mathbf{k}_t)=\mathbf{1}, ~t\in\{1,2\}$).

\begin{table}
 \centering
 \begin{tabular}{|c|c|c|c|c|c|c|}
 \hline
  TRP & 1 & 2 & 3 & 4 & 5 & 6 \\
 \hline
  TX & 1 & 2 & 3 & 3 & 5 & 4 \\
  \hline
  RX & 1 & 2 & 2 & 3 & 3 & 3 \\
  \hline
 \end{tabular}
 \caption{Look-up table mapping the TRP index with the corresponding TX and RX IDs, corresponding to Figs.~\ref{fig:pic_env2} and \ref{fig:lay_env2}.}
 \label{tab:lookup_scenario2}
\end{table}

The estimated target locations are plotted in Fig.~\ref{fig:res_env2}, from which the following inferences can be drawn:
\begin{figure}
 \includegraphics[scale=0.6]{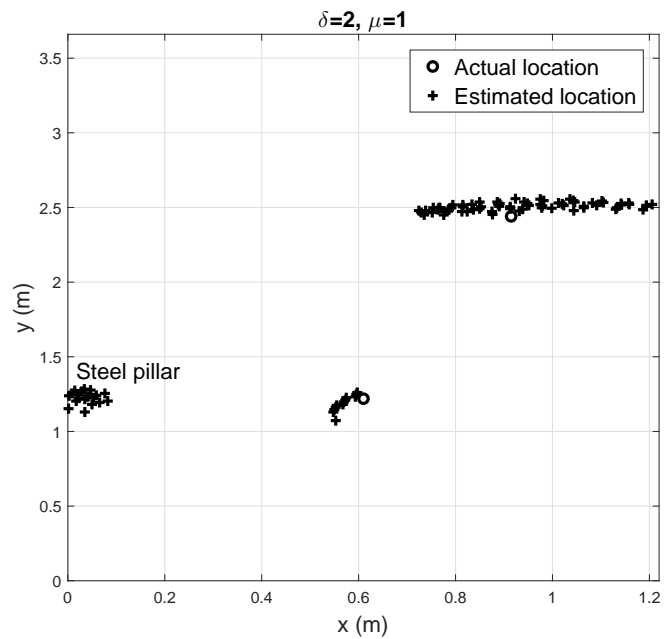}
 \caption{Position estimates for the targets obtained from the Bayesian MTL algorithm in the Measurement Scenario 2.}
 \label{fig:res_env2}
\end{figure}

\begin{itemize}
  \item[(i)] Setting $\mu=1$ ensures that only those points at which all six ellipses (corresponding to the six TRPs) intersect are detected as target locations. It can be seen that both targets are localized.
 \item[(ii)] The Bayesian MTL algorithm was formulated under the assumption of point targets. Since the targets are not point objects, multiple DPs are possible, in general, for each target. As a result, we obtain a \emph{cluster} of location estimates for each target location.
 \item[(iii)] The DPs to the targets are at least 10 dB above the noise floor post background cancellation, even in a cluttered environment, which can be attributed to the targets being strong reflectors and having a sufficiently large radar cross-section, due to the foil wrapping.
 \item[(iv)] The steel pillar to the left of Target 1, which is a part of the clutter, was still `localized' in spite of background subtraction. In terms of range, Target 1 and the pillar are closely separated for all the TRPs. Hence, some of the energy from the DP to Target 1 spills over into the delay bin corresponding to the pillar location. As a result, the pillar cannot be perfectly canceled out during background subtraction. The residual energy manifests itself as a DP to the pillar, leading to its localization. An implication of this is that the background in the immediate vicinity of a target cannot be subtracted completely.
\end{itemize}

\section{Summary and Conclusions}
\label{sec:summary}
In this paper, we considered the impact of environment-induced correlated blocking on localization performance. We first provided a theoretical framework for MTL using a distributed MIMO radar by formulating the general problem of localizing all the targets and scatterers in an unknown environment as a Bayesian estimation problem. We then proceeded to derive a more tractable approximation, known as the Bayesian MTL problem, where the objective was to localize only the targets, but not the scatterers. We then proposed a polynomial-time approximation algorithm - at the heart of which was a blocking-aware vector likelihood function that took correlated blocking into account - to solve this problem. The algorithm relies on two thresholds, $\delta$ and $\mu$, to detect targets and works with either theoretical or empirical blocking statistics that may be obtained via measurements or simulations. Our simulations showed that ignoring correlated blocking can be lead to very poor detection performance, with false alarms being more likely to occur than detections, and our experiments yielded encouraging results, even in the presence of non-idealities such as improper background subtraction and non-point targets.


\section{Acknowledgments}
The authors would like to thank C. Umit Bas, O. Sangodoyin and R. Wang for their assistance in carrying out the measurements.

\appendices
\section{Blocking model}
\begin{figure}
 \centering
 \includegraphics[scale=0.45]{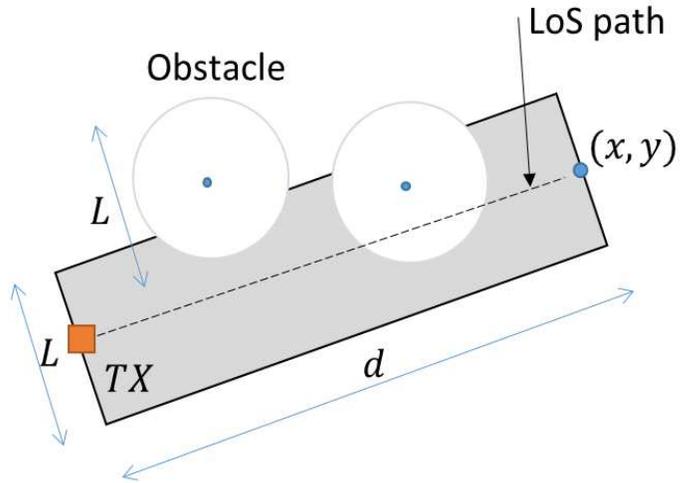}
 \caption{LoS is obstructed if there exists at least one scatterer center within a distance of $L/2$ from the LoS path}
 \label{fig:PPPbasics}
\end{figure}

Let the scatterers be represented by balls of diameter $L$, whose centers are distributed according to a homogeneous PPP with intensity $\lambda$. For LoS to exist between two points separated by a distance $d$, no scatterer center should lie within a rectangle of sides $L$ and $d$ (Fig. \ref{fig:PPPbasics}). Therefore, the LoS probability is $\exp(-\lambda L d)$.

\begin{figure}
 \centering
 \includegraphics[scale=0.28]{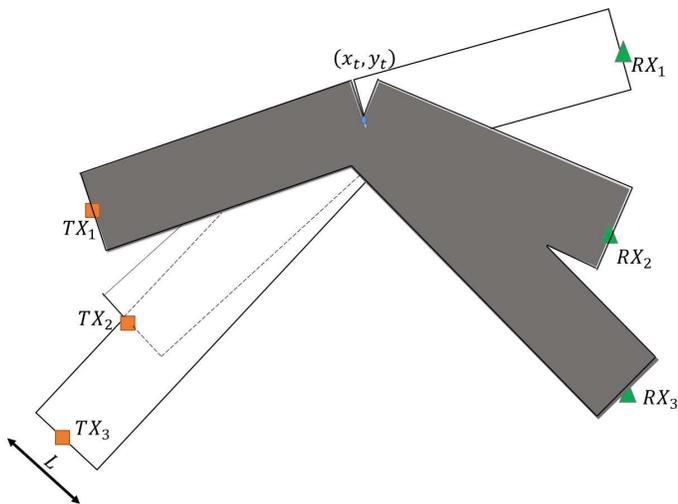}
 \caption{$\mathbf{k}_t=[0, 0, 0, 1, 0, 0, 1, 0, 0]=[0, 1, 1] \bigotimes [1, 0, 0]$}
 \label{fig:pmf}
\end{figure}

Consider a consistent blocking vector $\mathbf{k}_t=\mathbf{w}_t \bigotimes \mathbf{v}_t$ at $(x_t,y_t)$. The set of nodes that are blocked/unblocked at $(x_t,y_t)$ is determined by $\mathbf{v}_t$ and $\mathbf{w}_t$. For each unblocked node, there exists a rectangle which cannot contain any scatterer center. The \emph{LoS polygon}, $S_{\rm los}$, is the union of such rectangles (shaded grey in Fig. \ref{fig:pmf}). In contrast, for each blocked node $n$, there exists an NLoS polygon $S_n$ - the portion of its rectangle not contained in $S_{\rm los}$ - which must contain at least one scatterer center. Let $N_{\rm bl}$ denote the number of blocked nodes. Then,
\begin{align}
 \label{eq:prob_lbound}
 \mathbb{P}(\mathbf{k}_t) &\geq \exp(-\lambda {\rm Ar}(S_{\rm los})) \prod_{n=1}^{N_{\rm bl}}(1-\exp(-\lambda {\rm Ar}(S_n))) 
\end{align}
where ${\rm Ar}(.)$ denotes the area operator, acting on sets in $\mathbb{R}^2$. The expression in (\ref{eq:prob_lbound}) is a lower bound since it ignores overlapping NLoS polygons which may share scatterer centers (e.g., TX 2 and TX 3 in Fig. \ref{fig:pmf}). The bound is met with equality when none of the NLoS polygons overlap.
 
\label{app:blockmodel}

\bibliographystyle{IEEEtran}
\bibliography{IEEEabrv,JunyangBib}
\end{document}